\documentclass{article}
\usepackage{spconf,amsmath,graphicx}
\usepackage[acronym, nonumberlist]{glossaries}
\newacronym{6G}{6G}{Sixth Generation}
\newacronym{OFDM}{OFDM}{Orthogonal Frequency Division Multiplexing}
\newacronym{AI}{AI}{Artificial Intelligence}
\newacronym{DL}{DL}{Deep Learning}
\newacronym{BLER}{BLER}{Block Error Rate}
\newacronym{PTQ}{PTQ}{Post-Training Quantization}
\newacronym{QAT}{QAT}{Quantization-Aware Training}
\newacronym{SISO}{SISO}{Single-Input-Single-Output}
\newacronym{SIMO}{SIMO}{Single-Input-Multiple-Output}
\newacronym{MIMO}{MIMO}{Multiple-Input-Multiple-Output}
\newacronym{LLR}{LLR}{Log-Likelihood Ratio}
\newacronym{LoS}{LoS}{Line-of-Sight}
\newacronym{NLoS}{NLoS}{Non-LoS}
\newacronym{LS}{LS}{Least-Squares}
\newacronym{LMMSE}{LMMSE}{Linear Minimum Mean Squared Error}
\newacronym{ZF}{ZF}{Zero-Forcing}
\newacronym{CPU}{CPU}{Central Processing Unit}
\newacronym{GPU}{GPU}{Graphics Processing Unit}
\newacronym{NPU}{NPU}{Neural Processing Unit}
\newacronym{RAN}{RAN}{Radio Access Networks}
\newacronym{BER}{BER}{Bit Error Rate}
\newacronym{FLOP}{FLOP}{Floating Point Operation}
\newacronym{CDL}{CDL}{Clustered Delay Line}
\newacronym{LDPC}{LDPC}{Low-Density Parity-Check}
\newacronym{RG}{RG}{Resource Grid}
\newacronym{DMRS}{DMRS}{Demodulation Reference Signal}
\newacronym{IFFT}{IFFT}{Inverse Fast Fourier Transform}
\newacronym{FFT}{FFT}{Fast Fourier Transform}
\newacronym{BCE}{BCE}{Binary Cross-Entropy}
\newacronym{CSI}{CSI}{Channel State Information}
\newacronym{SNR}{SNR}{Signal-to-Noise-Ratio}
\newacronym{STE}{STE}{Straight‑Through Estimator}
\newacronym{UE}{UE}{User Equipment}
\newacronym{V2X}{V2X}{Vehicle-to-Everything}
\newacronym{QAM}{QAM}{Quadrature Amplitude Modulation}
\newacronym{PHY}{PHY}{Physical Layer}
\usepackage{amsmath,amssymb}
\usepackage{subcaption}
\usepackage{dsfont,comment,booktabs}
\usepackage[hidelinks]{hyperref}
\usepackage{xcolor}

\title{Efficient Quantization-Aware Neural Receivers: Beyond Post-Training Quantization}
\name{
  SaiKrishna Saketh Yellapragada$^{\star}$,
     Esa Ollila$^{\star}$,
  Mário Costa$^{\dagger}$
\thanks{The  work of  first author has been supported in parts by Research Council of Finland (grant no:359848) and European Union's 6GARROW project (No 101192194).}
}
\address{
  $^{\star}$Aalto University, Espoo, Finland \\
  $^{\dagger}$Nokia, Amadora, Portugal \\
  \texttt{\{saikrishna.yellapragada, esa.ollila\}@aalto.fi}, 
  \texttt{mario.costa@nokia.com}
}

\begin{document}
%
\maketitle

\begin{abstract}
As wireless communication systems advance toward \gls{6G} \gls{RAN}, \gls{DL}-based neural receivers are emerging as transformative solutions for \gls{PHY} processing, delivering superior \gls{BLER} performance compared to traditional model-based approaches. Practical deployment on resource-constrained hardware, however, requires efficient quantization to reduce latency, energy, and memory without sacrificing reliability. In this paper, we extend \gls{PTQ} by focusing on \gls{QAT}, which incorporates low-precision simulation during training for robustness at ultra-low bitwidths. In particular, we develop a \gls{QAT} methodology for a neural receiver architecture and benchmark it against a \gls{PTQ} approach across diverse 3GPP \gls{CDL} channel profiles under both \gls{LoS} and \gls{NLoS} conditions, with user velocities up to 40 m/s. Results show that 4-bit and 8-bit \gls{QAT} models achieve \glspl{BLER} comparable to \texttt{FP32} models at a 10\% target \gls{BLER}. Moreover, \gls{QAT} models succeed in \gls{NLoS} scenarios where \gls{PTQ} models fail to reach the 10\% \gls{BLER} target, while also yielding an $8\times$ compression. These results with respect to full-precision demonstrate that \gls{QAT} is a key enabler of low-complexity and latency-constrained inference at the \gls{PHY} layer, facilitating real-time processing in \gls{6G} edge devices.
\end{abstract}


%
%

\section{Introduction}
\label{sec:intro}
Deep Learning has recently delivered strong gains at the \gls{PHY}, where fully-convolutional neural receivers such as DeepRx replace channel estimation, equalization, and demapping to improve error-rate performance under standardized channels and mobility \cite{deeprx}. End-to-end designs further demonstrate that learned receivers can reduce or even eliminate pilot overhead while maintaining reliability, highlighting the practical promise of \gls{AI}-native \gls{PHY} stacks for beyond-5G and 6G systems \cite{e2e_faa}. However, deployment in edge-constrained radio hardware remains limited by compute, memory, and latency budgets, making precision reduction via quantization central to practical neural receiver inference. In this context, \gls{QAT} adapts the model during training to low-precision arithmetic \cite{quantization_whitepaper, wu2020integer,krishnamoorthi2018quantizing,jacob2017quantizationtrainingneuralnetworks, lsq+,lectures_songhan}.\par

{\gls{PTQ}} avoids retraining and compresses models by up to $8\times$, but can degrade at aggressive bit-widths and under distribution shifts characteristic of high‑Doppler channels. In prior work, a CNN‑based neural receiver with symmetric uniform {\gls{PTQ}} (per‑tensor and per‑channel) achieved near‑\texttt{float32} BLER at 8‑bit and competitive results at 4‑bit, establishing a strong baseline for efficient {\gls{PHY}} inference and motivating robustness analysis at low precision \cite{saketh_asilomar25,dfq,kuzmin2023pruning,simquant,yellapragada2025datafree}.
In this paper, we examine whether {\gls{QAT}} closes the remaining gap by training with simulated quantizers, thereby improving the dependability of ultra‑low‑bit neural receivers across mobility and channel conditions.

\gls{QAT} integrates simulated quantizers into the training graph so the network learns to compensate for rounding and range effects, and is widely shown to preserve accuracy at 8-bit and enable viable 4-bit operation across diverse architectures and hardware back ends. Concretely, {\gls{QAT}} co-designs scale and rounding behavior with task loss, retaining integer-arithmetic-only inference paths at deployment time to meet edge latency and energy targets without sacrificing model quality. These properties make \gls{QAT} a natural fit for neural receivers operating under time and frequency-selective fading, where amplitude spikes and distribution shifts can otherwise erode \gls{PTQ} accuracy at ultra-low bit-widths. \par

This paper advances efficient neural receivers beyond \gls{PTQ} by developing a \gls{QAT} pipeline for convolutional \gls{PHY} receivers. The principles discussed in this paper should apply to any convolutional-ResNet architectures. We evaluate on 3GPP CDL\mbox{-}B/D channels across low, medium, and high UE speeds using Sionna, a hardware-accelerated differentiable link-level setup \cite{sionna}. 

\paragraph*{Main contributions}
\begin{itemize}

  \item A principled \gls{QAT}  formulation with learnable clipping and per-channel scales, demonstrating that 4-bit \gls{QAT}  preserves deployment efficiency while maintaining accuracy.

  \item A comprehensive link‑level evaluation contrasting \gls{PTQ} and \gls{QAT} on 3GPP TR 38.901 CDL‑B/D channels across \gls{UE} speeds, with results reported at \gls{BLER} targets of 10\% and 1\%.

\end{itemize}

\section{Quantization-Aware Training} 
\label{sec:nrx_qat}

\gls{QAT} simulates low-precision inference during training by inserting fake-quantization operators in the forward pass, allowing the model to adapt to quantization noise and maintain higher accuracy after deployment. This allows for gradient-based optimization of both the model weights and the quantization parameters, despite the discrete nature of the quantization function. The process can be broken down into three distinct steps: clipping, quantization, and dequantization.

Let \(b\) denote the bit-width and \([\alpha, \beta]\) be the learnable clipping interval. For signed symmetric quantization, the integer range is $[q_{\min}, q_{\max}]$, where \(q_{\min}=-2^{b-1}\) and \(q_{\max}=2^{b-1}-1\).

\noindent \textbf{1. Clipping:} First, an input real-valued \(x\) is clipped to the learnable dynamic range:
\begin{equation}
\label{eq:clipping}
    x_c = \text{clamp}(x, \alpha, \beta) = \max(\alpha, \min(x, \beta)).
\end{equation}

\noindent \textbf{2. Quantization:} The clipped value is then quantized to an integer \(q\) using a uniform step size, or scale, $s$:
\begin{equation}
\label{eq:quantization}
    s = \frac{\beta-\alpha}{q_{\max}-q_{\min}}, \quad \text{and} \quad q = \Big\lfloor \frac{x_c}{s} \Big\rceil.
\end{equation}
For symmetric quantization, the zero-point is implicitly zero.

\noindent \textbf{3. Dequantization:} Finally, the integer \(q\) is mapped back to the real domain to produce the output of the fake-quantization operator, $F_b(\cdot)$:
\begin{equation}
\label{eq:dequant}
    F_b(x; \alpha, \beta) = s \cdot q.
\end{equation}
In the forward pass of \gls{QAT}, all activations and weights are replaced by the output of this operator. For the backward pass, the non-differentiable rounding function $\lfloor \cdot \rceil$ is handled using the \gls{STE} \cite{bengio2013estimating,understanding_STE}. This allows gradients to be propagated through the operator using the following approximations for the partial derivatives:
\begin{equation}
    \frac{\partial F_b}{\partial x} \approx \mathds{1}_{\{\alpha \le x \le \beta\}}, \quad
    \frac{\partial F_b}{\partial \alpha} \approx \mathds{1}_{\{x < \alpha\}}, \quad
    \frac{\partial F_b}{\partial \beta} \approx \mathds{1}_{\{x > \beta\}}.
\end{equation}
The gradient with respect to $x$ passes through unchanged only for values within the clipping range. The gradients for $\alpha$ and $\beta$ are non-zero only for clipped values, which serves to penalize the clipping bounds and encourages them to expand to fit the dynamic range of the inputs. This allows for the end-to-end training of both network and quantization parameters.

\subsection{Training Objective Under Simulated Quantization}
The training objective in \gls{QAT} is to minimize the task loss when model weights operate under quantized precision while maintaining full-precision activations. The neural receiver function 
$
f(\mathbf{s};W_q)
$
denotes the forward computation that processes an input signal $\mathbf{s}$ using quantized weights $W_q = F_b(W; \phi_w)$ to produce the network prediction, where $W$ denotes the network weights, $\phi_w$ represents the quantization parameters for weights (clipping bounds $\alpha$ and $\beta$), and $\mathcal{D}$ is the training data distribution with signal samples $\mathbf{s}$ and labels $\mathbf{y}$. Using the fake quantization operator $F_b(\cdot)$ defined in \eqref{eq:dequant}, the optimization problem becomes:
\begin{equation} \label{eq:qat_objective}
\min_{W, \phi_w} \mathbb{E}_{(\mathbf{y}, \mathbf{s}) \sim \mathcal{D}} \left[ \mathcal{L}\bigl(\mathbf{y},\, f(\mathbf{s};\, F_b(W; \phi_w))\bigr) \right]
\end{equation}
Note that in this work, activations remain in full precision throughout the network, allowing the {neural receiver} to preserve signal fidelity while achieving computational efficiency through weight quantization.

During quantization-aware fine tuning, weight updates are applied to full-precision master copies of $W$, ensuring proper gradient accumulation despite quantization noise. The quantization parameters $\phi_w$ are jointly optimized with the network weights to learn optimal dynamic ranges that minimize the quantization-aware loss.

\section{Quantization-Aware Neural Receiver} 

\subsection{System Model} \label{sec:system_model}
We consider an uplink \gls{SIMO} \gls{OFDM} system with $N_\text{Rx}$ receive antennas. The input bitstream is \gls{LDPC}-encoded, mapped to modulation symbols, and arranged into a \gls{RG} of size $N_\text{sym}\times N_\text{sc}$, with \glspl{DMRS} embedded at known time–frequency locations for channel estimation. After \gls{IFFT} and cyclic prefix insertion, the signal is transmitted over the 3GPP \gls{CDL} channel~\cite{3gpp2020channel}. At the receiver, \gls{FFT} yields
\begin{equation}
  \mathbf{y}_{n,k} = \mathbf{h}_{n,k}\, x_{n,k} + \mathbf{n}_{n,k},
  \quad \mathbf{y}_{n,k},\mathbf{h}_{n,k}\in \mathbb{C}^{N_\text{Rx}\times 1},
\end{equation}
where $x_{n,k}$ is the transmitted symbol, $\mathbf{n}_{n,k}\sim\mathcal{CN}(\mathbf{0},\sigma^2 \mathbf{I}_{N_\text{Rx}})$, and $\mathbb{E}[|x_{n,k}|^2]=1$. Channel estimates $\mathbf{h}_{n,k}$ at pilot positions are interpolated across the \gls{RG}. Traditionally, when the receiver is processing the \gls{OFDM} waveform, the post-\gls{FFT} waveform at the receiver is fed to perform channel estimation, equalization, and demapping.

\subsection{Neural receiver}

We consider a neural receiver architecture and the training strategy based on our previous work \cite{saketh_asilomar25}. This is designed to replace traditional signal processing operations where the input is the post-\gls{FFT} sequence and the output is \glspl{LLR}. The output of the network which are the \glspl{LLR}, are then used as input to \gls{LDPC} decoding.  The neural receiver is trained with the parameters mentioned in Table \ref{tab:training_parameters} using channel models \gls{CDL}-A, C, and E. This training scenario is a combination of \gls{NLoS} and \gls{LoS} models, while testing is performed using \gls{CDL}-B and \gls{CDL}-D.

\begin{table}[t]
\centering
\caption{Training Parameters and Randomization}
\label{tab:training_parameters}
\footnotesize
\begin{tabular}{lcc}
\toprule
\textbf{Parameter} & \textbf{Training} & \textbf{Randomization} \\
\midrule
Carrier Frequency     & 3.5 GHz                 & None \\
Channel Model         & CDL-[A,C,E]             & Uniform \\
RMS Delay Spread      & 10--100 ns              & Uniform \\
UE Speed           & 0--50 m/s               & Uniform \\
SNR                   & $-2$--15 dB             & Uniform \\
Subcarrier Spacing    & 30 kHz                  & None \\
Modulation Scheme     & 64-QAM                  & None \\
No. of Tx Antennas    & 1                       & None \\
No. of Rx Antennas    & 2                       & None \\
Code Rate             & 0.5                     & None \\
DMRS Configuration    & 3\textsuperscript{rd}, 12\textsuperscript{th} symbol & None \\
Optimizer             & Adam                    & None \\
Batch Size            & 128                     & None \\
\bottomrule
\end{tabular}
\end{table}

\begin{table}[t]
\centering
\caption{Testing Parameters}
\label{tab:test_parameters}
\footnotesize
\begin{tabular}{lc}
\toprule
\textbf{Parameter} & \textbf{Tested on} \\
\midrule
Channel Model & CDL-[B,D] \\
UE Speed &
\begin{tabular}[c]{@{}l@{}}
Low: 0--5.1 m/s \\
Medium: 10--20 m/s \\
High: 25--40 m/s
\end{tabular} \\
SNR & 0--12 dB \\
\bottomrule
\end{tabular}
\end{table}

The neural receiver 
is trained to maximize the bit-metric decoding performance \cite{bmdr_pavan}, 
using \gls{BCE} loss 
for quantifying the difference between the network's predicted \glspl{LLR} and the true coded bits 
across the complete \gls{OFDM} resource grid. The \gls{BCE} loss-function is 
\begingroup
\small
\begin{equation}
\mathcal{L}_{\mathrm{BCE}}(B, \hat{L}) =  {- }\mathbb{E}\left[B \log \sigma(\hat{L}) + (1 - B) \log (1 - \sigma(\hat{L}))\right],
\label{eq:batch_BCE_loss_training}
\end{equation}
\endgroup
where $B$ denotes the ground-truth coded transmitted bits, $\hat{L}$ denotes the predicted \glspl{LLR} and $\mathbb{E}[\cdot]$ expectation operator. The function $\sigma(\cdot)$ represents the sigmoid activation, which maps \glspl{LLR} to probabilities.

\subsection{QAT Fine-Tuning Protocol}

Building upon the baseline neural receiver trained with full-precision weights, we implement \gls{QAT} fine-tuning to recover performance degradation observed in our previous PTQ analysis. The \gls{QAT} process begins by initializing the fake quantization operators $F_b(\cdot)$ defined in \eqref{eq:dequant} throughout the neural receiver architecture. The quantization parameters $(\alpha, \beta)$ for each quantized layer are initialized based on the weight distributions of the full-precision baseline model to ensure stable convergence during fine-tuning.

The quantization-aware fine-tuning employs a conservative training regime designed to adapt the neural receiver to quantization noise while preserving the learned signal processing capabilities. After inserting simulated quantization operations into the neural receiver architecture, the model undergoes fine-tuning for up to 5000 epochs with a reduced learning rate of $10^{-6}$. This significantly lower learning rate compared to initial training prevents catastrophic disruption of the pre-trained representations while allowing gradual adaptation to quantization constraints.

The training objective remains the \gls{BCE} loss \eqref{eq:batch_BCE_loss_training}, enabling the model to jointly optimize both network weights $W$ and quantization parameters $\phi_w$ under the simulated low-precision regime. During each forward pass, all network weights are processed through their respective fake quantization operators, exposing the model to quantization noise patterns it will encounter during actual deployment.

\section{Experimental Results and Discussion}
\begin{figure*}[t]
    \centering
    \includegraphics[width=0.9\linewidth]{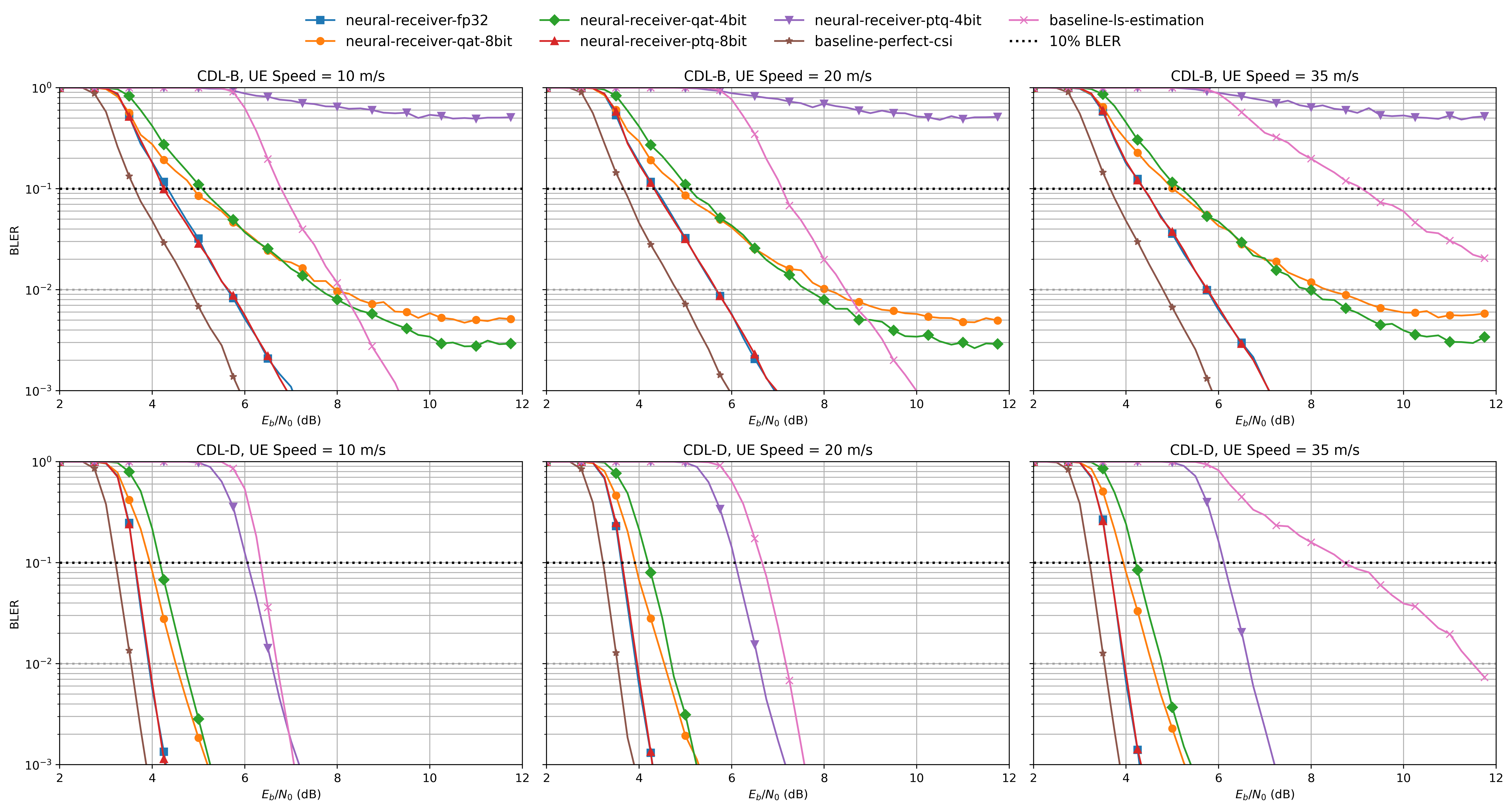}
    \caption{BLER vs. $E_b/N_0$ for the proposed neural receiver under CDL-B and CDL-D channels at various UE speeds.}

    \label{fig:BLER}
\end{figure*}

In this section, we evaluate the proposed neural receiver under $4$-bit and $8$-bit \gls{QAT} and \gls{PTQ}. Training is performed on \gls{CDL}-A, C, and E channels, with validation on \gls{CDL}-B and D models across the velocity ranges in Table~\ref{tab:test_parameters}, ensuring robust generalization. We report \gls{BLER} across varying \gls{UE} velocity, comparing QAT and PTQ receivers against a \texttt{FP32} neural receiver, \gls{LS}–\gls{LMMSE} equalization with soft demapping, and an ideal receiver with perfect \gls{CSI}.

{\bf NLoS comparisons.} The upper panel of Figure \ref{fig:BLER} illustrates the BLER performance under NLoS conditions,  using the CDL‑B channel model. The 8-bit \gls{QAT} configuration reaches 10\% BLER within an \gls{SNR} range of 4.88–5.01 dB across mobility scenarios, respectively, outperforming the \gls{LS} baseline by approximately 1.90–4.09 dB as mobility increases. At velocities ranging from 10 to 35 m/s, QAT 4‑bit requires 5.09–5.20 dB to attain 10\% BLER, maintaining a performance advantage over LS of approximately 1.71–3.90 dB, depending on the mobility level. Across 10–35 m/s, QAT‑8bit requires approximately 0.52–0.65 dB higher \gls{SNR} than \texttt{FP32} and PTQ‑8bit at 10\% BLER, whereas QAT‑4bit incurs roughly 0.73–0.80 dB over \texttt{FP32} and 0.75–0.84 dB over PTQ‑8bit, consistently across speeds.

{\bf LoS comparisons.}
The lower panel of Figure \ref{fig:BLER} illustrates the \gls{BLER} performance under \gls{LoS} conditions using the \gls{CDL}-D channel model. The 8-bit \gls{QAT} configuration achieves 10\% \gls{BLER} within an \gls{SNR} range of 3.94–3.97 dB across mobility scenarios, respectively, outperforming the \gls{LS} baseline by approximately 2.4–4.8 dB as mobility increases. At velocities ranging from 10 to 35 m/s, \gls{QAT} 4-bit requires 4.20–4.23 dB to attain 10\% \gls{BLER}, maintaining a performance advantage over \gls{LS} of approximately 2.19–4.50 dB, depending on the mobility level. Results show that quantization-aware training is essential for reliable performance at ultra-low bitwidths, as the 4-bit \gls{QAT} neural receiver consistently outperforms the \gls{PTQ} variant across all \gls{UE} speeds.

{\bf \gls{QAT} 8-bit vs \gls{QAT} 4-bit.}
The notably close performance between \gls{QAT} 8-bit and \gls{QAT} 4-bit, differing by only 0.18--0.27 dB in \gls{SNR} across channel conditions, demonstrates the effectiveness of QAT in learning robust low-precision representations. \gls{QAT} trains the model to be inherently resilient to quantization noise, reducing sensitivity to precision reduction from 8-bit to 4-bit. Additionally, during \gls{QAT}, the model learns weight distributions with fewer outliers and better-clustered values, facilitating aggressive quantization for neural network receivers. \par

{\bf \gls{QAT} 4-bit vs \gls{PTQ} 4-bit.}
A clear \gls{SNR} gap emerges between \gls{QAT} 4-bit and \gls{PTQ} 4-bit ($\sim$2–3 dB at 10\% \gls{BLER} under \gls{LoS}, with \gls{PTQ} failing to reach 10\% \gls{BLER} in \gls{NLoS}), underscoring the limitations of \gls{PTQ} at ultra-low bit-widths. \gls{QAT} adapts weight values during training to optimize for 4-bit representation, whereas \gls{PTQ} relies on pre-trained weights optimized for full-precision operation. In deep networks like the neural receiver, quantization errors accumulate across layers, which \gls{QAT} mitigates through training, unlike \gls{PTQ}, which cannot compensate for inter-layer error propagation. \par

Interestingly, at 8-bit precision, the \gls{PTQ} and full-precision \texttt{FP32} receivers exhibit a small \gls{SNR} advantage of $\sim$0.3 dB over the \gls{QAT} receiver. We attribute this to learned weight clipping in \gls{QAT}, which may under-represent extreme weights compared to \glspl{PTQ}’s calibration-free analytical rounding. Additionally, the \gls{STE}’s gradient bias and weight-grid oscillations during \gls{QAT} hinder convergence, leaving this residual gap that could affect radio performance.
\section{Conclusion and Future Work}
\label{conclusion}
The 4-bit \gls{QAT} meets 10\% \gls{BLER} targets in challenging \gls{NLoS} scenarios where 4-bit \gls{PTQ} fails, maintaining $\leq$0.8~dB \gls{SNR} gap to \texttt{FP32}. We intentionally kept channel and network configurations identical to \gls{PTQ} to enforce a controlled comparison and attribute gains specifically to the training methodology. Recent work shows that pruning and quantization trade off accuracy differently \cite{harma2025effective}, having established a \gls{QAT} baseline here, incorporating further compression methods via joint sparsity remains future work. Furthermore, our \gls{QAT} framework can be extended to low-bit activation quantization and mixed-precision weight–activation designs, which we also leave for future work.

The $8\times$ weight compression enables hardware-software codesign: on-chip SRAM and bandwidth can target 4-bit storage while maintaining near-\texttt{FP32} performance. This enables lightweight inference on resource-constrained edge devices, reducing memory and latency for \gls{6G} deployment. The strong \gls{LoS} performance of 4-bit \gls{QAT} receivers also positions them as viable candidates for integrated sensing and communication as well as site-specific deployments.
\vspace{-3mm}

\nocite{industryview6g,heresidual,kingma2014adam,4bitquant}
\newpage
\bibliographystyle{IEEEbib}
\bibliography{references}

\end{document}